\documentclass[sn-mathphys,Numbered]{sn-jnl}% Math and Physical Sciences Reference Style
\usepackage{graphicx}%
\usepackage{multirow}%
\usepackage{amsmath,amssymb,amsfonts}%
\usepackage{amsthm}%
\usepackage{mathrsfs}%
\usepackage[title]{appendix}%
\usepackage{xcolor}%
\usepackage{textcomp}%
\usepackage{manyfoot}%
\usepackage{booktabs}%
\usepackage{algorithm}%
\usepackage{algorithmicx}%
\usepackage{algpseudocode}%
\usepackage{listings}%

\usepackage{graphicx}
\usepackage{epsfig}
\usepackage{soul}

\usepackage{braket}
\usepackage{comment}
\usepackage{lipsum}
\usepackage{amssymb}
\usepackage{color}
\usepackage{mathtools}

\usepackage{amsmath}
 
\theoremstyle{thmstyleone}%
\theoremstyle{thmstyletwo}%

\theoremstyle{thmstylethree}%

\begin{document}

\title[Classical `spin' filtering with two degrees of freedom and dissipation]{Classical `spin' filtering with two degrees of freedom and dissipation}


\author[1,2]{\fnm{Atul} \sur{Varshney}}

\author[2]{\fnm{Areg} \sur{Ghazaryan}}

\author*[2]{\fnm{Artem} \sur{Volosniev}}\email{artem.volosniev@ist.ac.at}

\affil[1]{\orgdiv{School of Physical Sciences}, \orgname{National Institute of Science Education and Research, HBNI}, \orgaddress{\street{Jatni-752050}, \city{Odisha}, \country{India}}}

\affil[2]{\orgname{Institute of Science and Technology Austria}, \orgaddress{\street{am Campus 1}, \city{Klosterneuburg}, \postcode{3400}, \country{Austria}}}

\abstract{Coupling of angular motion to a spin degree of freedom gives rise to  various transport phenomena in quantum systems that are beyond the standard paradigms of classical physics.  
  Here, we discuss features of spin-orbit dynamics that can be visualized using a classical model with two coupled angular degrees of freedom. Specifically, we demonstrate classical `spin' filtering through our model and show that the interplay between angular degrees of freedom and dissipation can lead to asymmetric `spin' transport.  }



%\documentclass[aps,pra,twocolumn,groupedaddress]{revtex4-1}
%\usepackage{titlesec}

%\begin{document}

\maketitle
\section{Introduction}
Spin-orbit coupling (SOC) in classical and quantum systems is the interaction between spin and angular degrees of freedom. Many known phenomena can be understood using the concept of SOC. For example, synchronization of the Moon's spinning motion with its orbital motion$-$only one side of the Moon faces the Earth$-$is a result of spin-orbit coupling and energy dissipation in the Earth-Moon system~\cite{Darwin1879,Goldreich1966,barnes_tidal_2017}.
In condensed matter systems, the interaction between electron's spin and its angular momentum is crucial for the physics of spin-Hall effects~\cite{kato_observation_2004, sinova_universal_2004}, topological insulators~\cite{hasan_colloquium_2010}, spin textures in disordered systems \cite{koralek_emergence_2009}, spin-polarised current \cite{silsbee_spinorbit_2004}, to name a few. In optics, polarization of light plays the role of a spin degree of freedom, allowing one to observe similar phenomena~\cite{ozawa2019topological,bliokh2015spin}. 
 
It is also expected that SOC is key to recent observations 
of spin-polarised  photocurrents from substrates coated with self-assembled monolayers of chiral molecules (e.g. double-stranded DNA)~\cite{spin_2011, chiral-induced_2012}.
These observations introduced the concept of CISS (chiral-induced spin selectivity), which is triggering  interest in basic~\cite{Naaman2015,Evers2020,Evers2022} as well as in applied research~\cite{Naaman2015,Yang2021}. Certain aspects of CISS are still a subject of debate since the origin and strength of SOC in the problem is not known~\cite{Gersten2013, Gutierrez2013,Varela2016, Maslyuk2018,Dalum2019,Michaeli2019,Ghazaryan2020_filter,Ghazaryan2020,Liu2021}. State-of-the-art theoretical investigations are going beyond one-electron transport, and focusing on the role of the environment that leads to non-linear and non-unitary effects such as dephasing and energy dissipation~\cite{Guo2012, Li2020,Fransson2020,Zhang2020,Fransson2021,Alwan2021,Volosniev2021}. Studies of the latter are important beyond the CISS effect, as they contribute to a general understanding of the interplay between SOC and dissipation, which is now being explored in various physical systems \cite{Hata2021, Ghazaryan2022}. 
However, quantum analysis of dissipation is often intricate and we suggest in this work to also study relevant classical systems for building physical intuition for the SOC-dissipation interplay. 

\begin{figure}[t!]
\begin{center}
\includegraphics[width=7.5
cm]{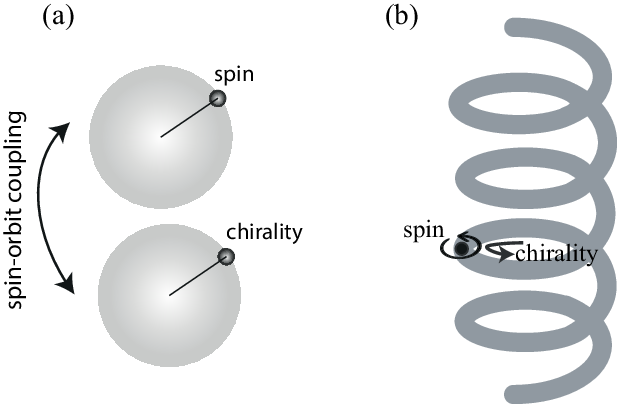}
\caption{(a) Illustration of a system
that we use to study the dynamics of  two coupled rotational motions.
The corresponding two classical degrees of freedom are called `spin', and `chirality' (or `orbit') in this paper, see the text for details. Correspondingly, their interaction is referred to as spin-orbit coupling (SOC). 
(b)   A chiral molecule in which an electron moves along the helical structure is a basic element in studies of the CISS. It provides a motivation for introducing the classical model in  (a).
} 
\label{fig1}
\end{center}
\end{figure}

In this paper, we introduce an arguably the simplest classical model where two rotational degrees of freedom\footnote{By `rotational degree of freedom' (also `angular degree of freedom' in this paper) we mean a degree of freedom that describes a planar pendulum in zero gravity or a particle in a ring.} are coupled in the presence of dissipation, see Fig.~\ref{fig1}. The model allows us to introduce the notions of `spin' and `chirality', and find conditions for classical `spin filtering' (analogous to CISS) in the weak SOC limit. 
One important remark is in order here. Electron’s spin is a quantum property that cannot be represented as a classical rotational motion. Therefore, our results do not explain the CISS effect, which we use to motivate the study. Our classical model only provides an illustrative example of basic spin-orbit dynamics that can lead to spin filtering effects, and can potentially be useful for constructing suitable quantum models in the future.

\section{Framework}

{\it Generalized spin-orbit coupling.} To study the coupling between two classical rotational degrees of freedom, we  consider the basic time-reversal and rotationally symmetric Lagrangian:
\begin{equation}
    L=\beta_1 \dot \theta_1^2 + \beta_2 \dot \theta_2^2 + \gamma \dot \theta_1 \dot\theta_2 f(\theta_2-\theta_1),
    \label{eq:Lagrang}
\end{equation}
where $\theta_1$ ($\theta_2$) is the `orbital' (`spin') degree of freedom; $\beta_1>0$, $\beta_2>0$ and $\gamma>0$ are parameters of the system. $\theta_1$ and $\theta_2$ are classical variables defined for convenience on the real axis, i.e., $\theta_i\in (-\infty,\infty)$. Physically the system returns to itself if $\theta_i\to\theta_i+2\pi$. The function $f$ determines the SOC  and depends only on the difference between $\theta_1$ and $\theta_2$, ensuring that the total angular momentum of the system is conserved (see App.~\ref{app:general_solution}). It is also periodic $f(x)=f(x+2\pi)$.

Without loss of generality, we set $\beta_1=1$, $\beta_2=\beta$ and assume that the maximal value of $|f|$ is unity. Our focus is on the case of weak SOC ($\gamma\to0$). This limit allows us to treat `orbit' and `spin' as separate well-defined quantities because the energy exchange between these degrees of freedom is small. Weak SOC is also relevant for CISS, as organic molecules consist of light atoms (carbon, hydrogen, oxygen, etc.).

{\it Chirality and the spin projection.} We say that the system has left [right] chirality if $\dot \theta_1(t=0)>0$ [$\dot \theta_1(t=0)<0$]. This assumption is motivated by the tight-binding representation of a helical molecule, in which the electron moves in one spatial dimension~\cite{Guo2012}.
Naturally, the sign of $\dot \theta_1$ can lead to non-trivial effects only in the presence of SOC, i.e., for $\gamma\neq0$.
In general, the sign of $\dot \theta_1$ is not a conserved quantity, however, for the systems we consider below (weak SOC), this will be the case. 

By analogy, we say that $\dot \theta_2>0$ [$\dot \theta_2<0$] corresponds to `spin-up' and `spin-down' particles. Our work shall illustrate that `spin filtering' in principle does not require quantum nature of particles. In particular, `spin filtering'  is possible with our classical interpretation of spins. We thus note  that
a deeper understanding of the role of quantum physics in the CISS effect is required for developing CISS-based quantum technologies~\cite{Aiello2022}.

Note that a simultaneous change of signs of $\dot \theta_1$ and $\dot \theta_2$ does not lead to any qualitative change, which is a manifestation of time-reversal symmetry. Therefore, without loss of generality we shall consider $\dot \theta_1>0$ and change the sign of $\dot \theta_2$ in the analysis of the system.

{\it Dissipation.} To include dissipation into the system, we rely on the Rayleigh dissipation function, $G$, chosen from empirical 
considerations~\cite{Goldstein1980}. In our case, we assume that frictional forces are proportional to velocities so that $G=(\alpha_1\dot \theta_1^2+\alpha_2 \dot \theta_2^2)/2$ ($\alpha_i>0$), which leads to the equations of motion
\begin{equation}
    \frac{\partial L}{\partial \theta_i}-\alpha_i\dot \theta_i=\frac{\mathrm{d}}{\mathrm{d}t}\frac{\partial L}{\partial \dot\theta_i}.
\end{equation}
 We shall assume that there is no dissipation associated with the `spin' degree of freedom $\alpha_2=0$, i.e., we assume `spin-conserving' interactions with the bath that causes dissipation, and write $\alpha=\alpha_1$ for simplicity. It is worth noting that, for the CISS effect, such an assumption is inherently reasonable, given long spin-relaxation times in organic molecules \cite{Dediu2009}.

\section{Angle-independent SOC}
\label{sec:angle_ind_SOC}
 To provide some physical intuition into the problem,
 let us consider the simplest situation: $f=1$. It corresponds to the position-independent SOC, typical for condensed matter systems.
 The case for $f=1$  is also the closest analogue of the phenomenological quantum model of Ref.~\cite{Volosniev2021}, which focuses on the interplay between spin-orbit coupling and dissipation in the context of the CISS effect.

Before we proceed, we note that the simplest physical realization of Eq.~(\ref{eq:Lagrang}) with $f=1$ corresponds to two uncoupled rotational degrees of freedom whose Lagrangian,
\begin{equation}
L=\dot \Theta_1^2+\dot \Theta_2^2,
\label{eq:lagrangian_two_theta}
\end{equation}
upon the transformation $\Theta_1=\theta_1+\theta_2$ and $\Theta_2=\theta_2$ has the form of Eq.~(\ref{eq:Lagrang}) with $\beta=2$ and $\gamma=2$. Dynamics of this system becomes non-trivial if we include dissipation coupled to $\dot \theta_1$. Physically, this implies that the $\Theta_2$ degree of freedom defines a rest frame for the $\Theta_1$ degree of freedom.

For a general form of Eq.~(\ref{eq:Lagrang}), we derive time evolution of $\theta_1$ and $\theta_2$ degrees of freedom
\begin{align}
    \theta_1=\theta_1^0+\dot \theta_1^0\frac{C}{2\alpha}\left(1-e^{-\frac{2\alpha t}{C}}\right), \\ \theta_2=\theta_2^0+\left[\frac{\gamma}{2\beta}\dot\theta_1^0+\dot\theta_2^0\right]t-\frac{\gamma C}{4\beta\alpha}\dot \theta_1^0\left(1-e^{-\frac{2\alpha t}{C}}\right),
\end{align}
where $C=4-\gamma^2/\beta$\footnote{Note that at the special point $\beta=\gamma^2/2$ the system without dissipation can be parameterized by a single degree of freedom $\theta_1+\theta_2$.}, and superscript $0$ means that the function should be taken at $t=0$, e.g., $\theta_1^0=\theta_1(t=0)$.
Note that the $\theta_2$-pendulum learns about the initial state of the $\theta_1$-pendulum only if $\gamma$ and $\alpha$ are non-vanishing\footnote{Note that if $\alpha\to0$ then $(1-e^{-2\alpha t/C})/\alpha\to 2t/C$.}.
This demonstrates the necessity to have 
both the SOC and dissipation for classical `spin filtering' discussed below.

At $t\to\infty$, the orbital degree of freedom reaches the value $\theta_1^f=\theta_1^0+\dot\theta_1^0 C/(2\alpha)$, and then its dynamics stops. Using the picture of the CISS effect, this value can be interpreted as the distance an electron travels in a molecule before it loses all of its energy associated with the orbital degree of freedom. To investigate $\theta_1^f$, we assume that the initial energy, $E$, and $|\dot \theta_2^0|$ (i.e., the `spin' degree of freedom) are fixed. 
We derive
\begin{equation}
\theta_1^f=\theta_1^0+\frac{C}{2\alpha}\frac{\sqrt{\gamma^2(\dot\theta_2^0)^2+4(E-\beta(\dot \theta_2^0)^2)}-\gamma\dot\theta_2^0}{2}.
\end{equation}
The key observation here is that by changing the sign of $\dot\theta_2$ we change the distance $\theta_1^f$. This can lead to `spin filtering', as the `spin-down' $(\dot \theta_2^0<0)$ can travel farther than the `spin-up'   $(\dot \theta_2^0>0)$: $\Delta\theta_1^f=C\gamma |\dot \theta_2^0|/(2\alpha)$.
To interpret this `spin filtering' in the language of the CISS effect, note that if the total electron path in the molecule is longer than $l\theta_1^{f}(\dot \theta_2^0>0)$ but shorter than $l\theta_1^{f}(\dot \theta_2^0<0)$ then only spin-down electrons can go through the molecule ($l$ sets the unit of length here).

One of the features of the `spin-filtering' process for $f=1$ is that (just like for systems with quantum SOC) the value of $\dot\theta_1^0$ must be changed together with `spin' orientation to fix the total energy. As we demonstrate in the following sections other classical SOC couplings can have $f(\theta_2-\theta_1)=0$ at $t=0$, which is beyond the typical setting in condensed matter physics.  Such couplings allow us to investigate systems with the same value of $\dot \theta_1^0$ for different `spins'.

Finally, we interpret results of this section in terms of $\Theta_1$ and $\Theta_2$, see Eq.~(\ref{eq:lagrangian_two_theta}). In this case,
$\dot \theta_1^0=\dot \Theta_1^0-\dot \Theta_2^0$. By fixing $E$ and $|\dot \Theta_2^0|$, we also fix $|\dot \Theta_1^0|$. 
The change of sign $\dot \Theta_2^0$ then naturally leads to a change in the value of $\dot \theta_1^0$, which determines $\theta_1^f$ and hence `spin filtering' properties of the system.

\begin{figure}[t!]
\begin{center}
\includegraphics[width=4cm]{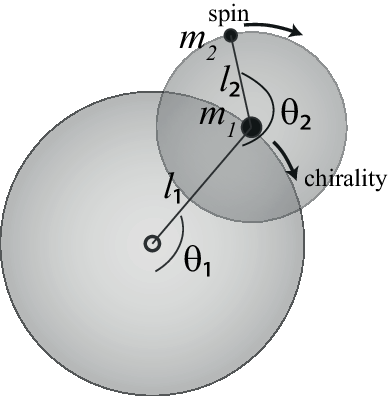}
\caption{Double pendulum that consists of two masses $m_1$ and $m_2$ attached to two massless rods of length $l_1$ and $l_2$. We interpret the motion of the mass $m_2$ as `spin' dynamics, and the direction of the motion of the mass $m_1$ as `chirality'. } 
\label{fig2}
\end{center}
\end{figure}

\section{Double pendulum}
Another physical system described by the Lagrangian from Eq.~(\ref{eq:Lagrang}) is a planar double pendulum without gravity. We use the standard parametrization of this textbook system: two point masses $m_1$ and $m_2$ attached to massless rods of fixed lengths $l_1$ and $l_2$, see Fig.~\ref{fig2}(a), to write  
\begin{equation}
L=\frac{M l_1^2}{2}\dot{\theta}_1^2+\frac{m_2l_2^2}{2}\dot{\theta}_2^2+m_2l_1l_2\dot{\theta}_1\dot{\theta}_2\cos(\theta_1-\theta_2),
\end{equation}
where $M=m_1+m_2$. Note that $f=\cos(\theta_1-\theta_2)$, meaning the SOC here depends on the relative angle unlike the system in the previous section. Before we proceed we note that a planar double pendulum without gravity has been studied extensively in classical mechanics, see, e.g.,~\cite{Enolskii2003,Szuminski2014}, however (to the best of our knowledge) not in the context of the present work.
To add dissipation, we use the following Rayleigh function
$G=\alpha l_1\dot \theta_1^2/4$, where the strength of dissipation is parameterized by $\alpha$.

To find time dynamics of a dissipative double pendulum, we solve the Lagrange's equations numerically using a finite-difference method, see App.~\ref{app:numerics}.
One remark is in order here. The double pendulum in a gravitational field $-$the standard textbook example of a double pendulum$-$ is a chaotic system \cite{Shinbrot1992} sensitive to small perturbations, which can amplify numerical errors.
In contrast, the double pendulum in zero gravity is an integrable system, see, e.g.,~\cite{Enolskii2003}, making our calculations less prone to numerical errors.

\begin{figure}[t!]
\begin{center}
\includegraphics[width=8.5cm]{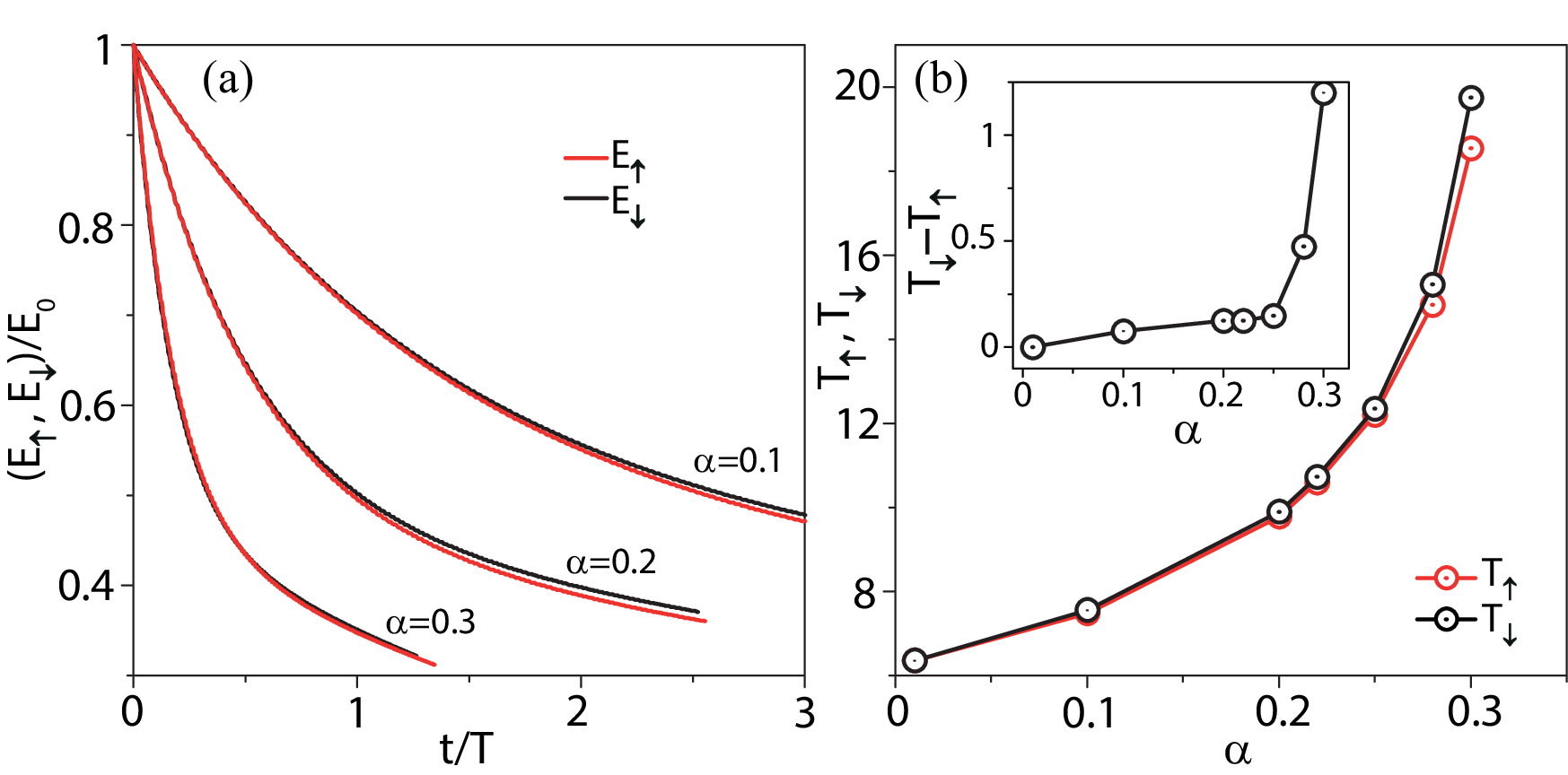}
\caption{ (a)   Energy decay  of the system, normalized with the initial  energy $E_0$, as a function of  $t/T$ for spin-up (red) and spin-down (black) systems at various $\alpha$. The initial conditions are $\theta_1^0-\theta_2^0=\pi /4$, $|\dot\theta_1^0|=1$, and $|\dot\theta_2^0|=30$.
Our units are chosen such that $l_1=1$ and $M=1$. In these units, we take $m_2=0.1$, $m_1=0.9$, and $l_2=0.1$.  (b) Orbital period $T$ as a function of $\alpha$ for `spin-up' (red) and `spin-down' (black). The initial conditions are as in (a).    Inset: Difference in the orbital periods of `spin-down' and `spin-up', i.e., $T_{\downarrow}-T_{\uparrow}$,  versus $\alpha$.} 
\label{fig3}
\end{center}
\end{figure}

To illustrate time evolution, we compute the instantaneous energy of the system, $E(t)=L$,  for `spin-up' and `spin-down' dynamics. As we have many parameters, we rely on the following strategy to choose initial conditions. 
First, we calculate trajectories of a non-dissipative system with 
$\theta^0_1-\theta^0_2=\pi/2$ and
$|\dot\theta_i^{0,\uparrow}|=|\dot \theta_i^{0,\downarrow}|$. This choice of initial velocities is possible as the spin-orbit interaction term vanishes at $t=0$  [here, $f(\pi/2)=0$].
 The energy of the system is then `spin'-independent: $E_{\uparrow}=E_{\downarrow}$. Second, we choose the initial conditions from this trajectory by fixing $\theta^0_1-\theta^0_2$ and finding the corresponding velocities. 
By choosing $\theta_i(t=0)$ and $\dot \theta_i(t=0)$ in this way, we explore the dynamics of the system for different initial conditions, but for the same value of the initial energy,~$E_0$. 

Figure \ref{fig3}(a) shows time evolution of $E_{\uparrow}$ and $E_{\downarrow}$ for the initial conditions that correspond to  $\theta^0_1-\theta_2^0=\pi/4$ at $t=0$. To present this evolution in dimensionless form, we use the period of `orbital' motion, $T$, see the next section for a formal definition.
Note that the dissipation of energy is almost independent of the spin. We observed such a behavior for various initial conditions and parameters of the system. To understand this, note that the sign of spin-orbit coupling is  ill-defined for this system. Indeed, the function $f$ changes its sign during time evolution. As we explain in the next section, this is the reason there can be no efficient `spin filtering'. Figure \ref{fig3}(b) shows the period of `orbital' motion for different strength of dissipation. Note that by increasing the value of $\alpha$, one can force the period to strongly depend on `spin'. However, this can be used for `spin-filtering' only if one is able to fix the initial conditions -- other initial conditions would favor another `spin'.

\section{General form of SOC}

To study time dynamics with a general form of SOC, we first discuss the system without dissipation and show that the period of `orbital' motion depends on both `spin' and `chirality'. Then, we connect the calculated period to $\theta_1^f$, which is a suitable measure of classical `spin filtering', see Sec.~\ref{sec:angle_ind_SOC}.
Note that the system is integrable (see App.~\ref{app:general_solution}) for all values of $\gamma$. We focus only on weak SOC, which  
provides the most clear physical picture of the dynamics.

{\it Weak SOC without dissipation.} In the limit of weak SOC, i.e., $\gamma\to 0$, the effect of SOC can be treated perturbatively. 
We derive
\begin{align}
\dot \theta_1\simeq \dot\theta_1^0-\frac{\gamma (\dot\theta^0_2)^2(f(x)-f(\delta))}{2(\dot\theta^0_2-\dot\theta^0_1)},\\
\dot \theta_2\simeq \dot\theta_2^0+\frac{\gamma (\dot\theta^0_1)^2  (f(x)-f(\delta))}{2\beta(\dot\theta^0_2-\dot\theta^0_1)},
\label{eq:weak_SOC_no_dissip}
\end{align}
where $x=\dot\theta^0_2 t-\dot\theta^0_1 t + \delta$, $\delta=\theta_2(t=0)-\theta_1(t=0)$, see App.~\ref{app:weak_SOC_no_dissipation}.

Let us now calculate the period $T$ of `orbital' motion, defined via $\int_0^T\dot \theta_1\mathrm{d}t=2\pi$:
\begin{equation}
\left(\frac{T}{2\pi}\right)^{-1}\simeq \dot\theta^0_1 - \frac{\gamma (\dot\theta^0_2)^2}{2(\dot\theta^0_2-\dot\theta^0_1)}\langle f\rangle,
\label{eq:period_main}
\end{equation}
where $\langle f\rangle=\int_0^T (f(x)-f(\delta))\mathrm{d}t/T$. 
We see that the period for $\dot\theta^0_2<0$ is different from the period with $\dot\theta^0_2>0$ if $\langle f\rangle \neq 0$. 
We fix $\dot \theta_1^0$, and write the difference in $T$ between systems with `spin-up' ($\dot\theta^0_2>0$) and `spin-down' ($\dot\theta^0_2<0$) motions as
\begin{equation}
\Delta T\simeq \frac{2\pi \gamma (\dot \theta_2^0)^2 \langle f\rangle}{\dot \theta_1^0 ((\dot \theta_2^0)^2-(\dot \theta_1^0)^2)}.
\label{eq:deltaT}
\end{equation}
This difference depends on the strength of SOC and on the chirality, i.e., the sign of $\dot \theta_1^0$, as expected. 
Note that $\Delta T$ vanishes if $\langle f\rangle=0$ as, e.g., for $f(x)=\cos(x)$.
This happens because the SOC changes its sign during time evolution. In other words, the sign of $\gamma$ is not well-defined for an oscillating function $f$ with equal contributions of positive and negative parts to $\langle f\rangle$. 

{\it Weak SOC with dissipation.}
The dissipative system with weak SOC also allows us to derive an approximate solution, see App.~\ref{app:weak_SOC_with_dissipation}. As we are interested in spin-filtering properties of the system, we calculate the quantity $\theta_1^f$ 
\begin{equation}
\theta_1^f=\theta_1^0 +\frac{2\dot \theta_1^0}{\alpha}-\lim_{z\to\infty}\int_0^z\frac{\gamma (\dot \theta_2^0)^2 (f(x)-f(\delta))}{2(\theta_2^0-\theta_1^0) }e^{\frac{\alpha (t-z)}{2}}\mathrm{d}t,
\end{equation}
where $x$ and $\delta$ are defined after Eq.~(\ref{eq:weak_SOC_no_dissip}).
Note that this expression reproduces the result of Sec.~\ref{sec:angle_ind_SOC}
for $f=1$. 

We see that if `spin' motion is fast in comparison to all other timescales of the problem, e.g., rate of dissipation and `orbital' motion, then we can use average values in place of $f(x)$. In this case the period $T$ from Eq.~(\ref{eq:period_main}) determines $\theta_1^f$:  $\theta_1^f\simeq \theta_1^0+4\pi/(\alpha T)$. The difference in periods of `orbital' motion for `spin up' and `spin down' then leads to classical `spin filtering'. 
According to Eq.~(\ref{eq:deltaT}), the `spin filtering' effect occurs 
even if $\dot \theta_i^{0,\uparrow}=\dot \theta_i^{0,\downarrow}$ as long as $\langle f\rangle \neq0$. This case can be realized with (for example) $f(x)\sim \cos^2(x)$ that can enjoy $f(\delta)=0$. Although, this type of SOC does not have an obvious analogue in condensed matter physics, it clearly shows that the interplay between `spin' and `chirality' can lead to `spin filtering' even for identical initial conditions.

\section{Conclusions} 

In conclusion, we examined a classical dissipative system involving two coupled rotational degrees of freedom.  We identified two scenarios with respect to the form of spin-orbit coupling, $f$: (i) $\langle f\rangle\neq 0$, e.g. $f=1$; (ii)  $\langle f\rangle = 0$, e.g. $f(x)=\cos(x)$. We demonstrated that the former case exhibits `spin-filtering' effect that depends on both the `chirality' and `spin' properties. It remains a possibility that `spin-filtering' effects may manifest in the latter case under specific initial conditions (see Fig.~\ref{fig3}), however, this cannot be generalized universally.

 We highlighted some similarities between our set-up and models of the CISS effect.  It is essential to acknowledge that classical rotational motion is not representative of an electron's spin. Consequently, our framework cannot provide an exhaustive explanation of the observed CISS effect.  In addition, it is important to note that our current model omits consideration of the substrate, a factor that could potentially have significant implications in elucidating experimental outcomes~\cite{Gersten2013,Liu2021,fransson-nano-2021,Adhikari2023,Alhyder2023}.
Nevertheless, in the context of CISS, our simple set-up provides a rudimentary model that can be used for illustrative purposes. Note a recent work~\cite{Chen2023} where another classical analogue of the CISS effect is introduced for a similar purpose. However, unlike the present work,  Ref.~\cite{Chen2023} uses a complex system where a charged particle moves in a helical dissipative environment, and where  spin-orbit coupling is being generated by friction.

In the paper, we focused on the case when the `orbital' and `spin' degrees of freedom are defined in the same plane, which naturally corresponds to $L_z\sigma_z$-type of spin-orbit coupling in quantum physics. It is easy to design classical analogues of other cases, e.g., $k_z\sigma_z$-type of SOC may loosely correspond to a double pendulum without gravity with the `spin' motion in a plane perpendicular to the `orbital' motion. Subsequent investigations could explore the `spin' and `orbit' motions in different planes,  such an approach may unveil phenomena not addressed in the current study, 
 for instance, the emergence of classical geometric (Berry) phases.

\bmhead{Acknowledgments}

We thank Mikhail Lemeshko and members of his group for many inspiring discussions; Alberto Cappellaro for comments on the manuscript.

\newpage

\begin{appendices}

\section{Equations of motion for a double pendulum}
\label{app:numerics}

The Lagrangian for a double pendulum without dissipation reads as
\begin{equation}
L=\frac{1}{2}m_1l_1^2\dot{\theta}_1^2+\frac{1}{2}m_2\left[l_1^2\dot{\theta}_1^2+l_2^2\dot{\theta}_2^2+2l_1l_2\dot{\theta}_1\dot{\theta}_2\cos(\theta_1-\theta_2)\right].
\end{equation}
To add dissipation to our model, we use the following function
$G=(l_1\alpha\dot \theta_1^2+l_2\alpha_s \dot \theta_2^2)/4$, where for the sake of generality we have assumed that both angular degrees of freedom are subject to dissipation. 
With this form, we derive the following equations of motion
\begin{align*}
    \frac{\partial^2 \theta_1}{\partial t^2}=\frac{-m_2l_1\dot \theta_1^2\sin(2\theta_1-2\theta_2)-2m_2l_2\dot\theta_2^2\sin(\theta_1-\theta_2)+\alpha_s\dot \theta_2\cos(\theta_1-\theta_2)-\alpha\dot \theta_1}{l_1(2m_1+m_2-m_2\cos(2\theta_1-2\theta_2))},\\
    \frac{\partial^2 \theta_2}{\partial t^2}=\frac{m_2l_2\dot \theta_2^2\sin(2\theta_1-2\theta_2)+2l_1(m_1+m_2)\dot\theta_1^2\sin(\theta_1-\theta_2)-\alpha_s \frac{(m_1+m_2)}{m_2}\dot \theta_2+\alpha\dot \theta_1\cos(\theta_1-\theta_2)}{l_2(2m_1+m_2-m_2\cos(2\theta_1-2\theta_2))}.
    \end{align*}
We re-write these equations as a system of first-order differential equations by introducing $v_i=\dot \theta_i$, and use the Runge-Kutta integration method to find a numerical solution. We benchmarked against analytically exact results available for $\alpha=\alpha_s=0$ (see below) to investigate convergence of our results.

  \section{General solution to the system without dissipation.} 
  \label{app:general_solution}

Here, we work in a co-moving frame defined by a set of variables
$\phi_1=\theta_1$ and $\phi_2=\theta_2-\theta_1$; the corresponding Lagrangian is independent of $\phi_1$: $
    L=\dot \phi_1^2 + \beta (\dot \phi_1+\dot \phi_2)^2 + \gamma \dot \phi_1 (\dot\phi_1+\dot\phi_2) f(\phi_2)$.
Following the standard approach of classical mechanics, we introduce the generalized momenta $p_i=\partial L/\partial \dot \phi_i$
\begin{equation}
\begin{aligned}
    p_1=2 \dot \phi_1+2\beta(\dot\phi_1+\dot\phi_2) + \gamma(2\dot\phi_1+\dot\phi_2)f(\phi_2),\\ p_2=2 \beta (\dot \phi_1+\dot \phi_2)+\gamma\dot \phi_1 f(\phi_2),
    \end{aligned}
\end{equation}
  and study the system in the Hamiltonian formalism\footnote{This Hamiltonian can also be written as
  $H=(p_1\dot \phi_1 + p_2\dot\phi_2)/2$,
where $
    \dot \phi_2=(2 p_2-\gamma p_1 f(\phi_2)+2\gamma p_2 f(\phi_2)-2\beta p_1+2\beta p_2)/(4\beta- \gamma^2 f(\phi_2)^2)$ and
$\dot \phi_1=(2\beta p_1-2\beta p_2-\gamma p_2f(\phi_2))/(4\beta-\gamma^2 f(\phi_2)^2)$}:
\begin{equation}
\begin{aligned}
     H= \frac{p_1^2 + A(\phi_2)p_2^2-2 B(\phi_2)p_1 p_2}{C(\phi_2)},
    \end{aligned}
    \label{eq:Hamiltonian}
\end{equation}
where $A(\phi_2)=(1+\gamma f(\phi_2)+\beta)/\beta$, $B(\phi_2)=(\gamma f(\phi_2)+2\beta)/(2\beta)$, and $C(\phi_2)=4-\gamma^2 f(\phi_2)^2/\beta$. As a result of rotational invariance, there is no dependence on $\phi_1$ and thus the total angular momentum, $p_1$, is conserved. The dynamics of the system effectively corresponds to one-body motion parameterized by $p_2,\phi_2$.

The Hamiltonian $H$ does not depend on time $t$  explicitly -- the energy is conserved $H\to E$.  To find other integrals of motion, we use the equation
\begin{equation}
    \dot \phi_2=\frac{\partial H}{\partial p_2} \to \dot \phi_2 = \frac{2}{C(\phi_2)}\left(A(\phi_2)p_2-B(\phi_2)p_1\right),
\end{equation}
where the value of $p_2$ (for a given $\phi_2$) can be determined from the energy constraint:
\begin{equation}
    \mathcal{P}_2(\phi_2)=\frac{B(\phi_2)\mathcal{P}_1}{A(\phi_2)}\pm \sqrt{\left(\frac{B(\phi_2)\mathcal{P}_1}{A(\phi_2)}\right)^2-\frac{\mathcal{P}_1^2-C(\phi_2)E}{A(\phi_2)}},
\end{equation}
where we have used $\mathcal{P}_1=p_1$
and $\mathcal{P}_2(\phi_2)=p_2$ to emphasize that $p_1$ is a conserved quantity, and $p_2$ depends on $\phi_2$. The sign in the equation is determined by the initial condition. 
For $\dot\phi_2\neq0$\footnote{If $\dot\phi_2=0$ at some $t_{\dot\phi_2=0}$, one should consider time evolution on the intervals with $t \lessgtr t_{\dot\phi_2=0}$ separately and then smoothly connect the resulting dynamics at $t_{\dot\phi_2=0}$.}, we calculate $t(\phi_2)$
\begin{equation}
    t=\int \frac{C(\phi_2) \mathrm{d}\phi_2}{2(A(\phi_2)\mathcal{P}_2(\phi_2)-B(\phi_2)\mathcal{P}_1)} + \mathrm{const}_1,
    \label{eq:t}
\end{equation}
where $\mathrm{const}_1$ is an integral of motion. Without loss of generality, we can always redefine the origin of time-axis such that $\mathrm{const}_1=0$. 
Now, one can also easily find $\phi_1(\phi_2)$:
\begin{equation}
    \phi_1=\int\frac{\mathcal{P}_1-B(\phi_2)\mathcal{P}_2(\phi_2)}{A(\phi_2)\mathcal{P}_2(\phi_2)-B(\phi_2)\mathcal{P}_1}\mathrm{d}\phi_2 + \mathrm{const}_2,
    \label{eq:phi_1}
\end{equation}
where $\mathrm{const}_2$ is the last integral of motion. One can get rid of it by properly choosing the system of coordinates. 

The derived equations provide a complete picture of the effects of spin-orbit coupling in the introduced system. To study this effect, it is convenient to introduce the phase $\phi_B$\footnote{Note that this phase can be negative depending on the direction of the `spin'.} acquired by the $\phi_2$-pendulum after one period of the $\phi_1$-pendulum
\begin{equation}
2\pi=\int_{\phi_2(t=0)}^{\phi_2(t=0)+\phi_B}\frac{\mathcal{P}_1-B(\phi_2)\mathcal{P}_2(\phi_2)}{A(\phi_2)\mathcal{P}_2(\phi_2)-B(\phi_2)\mathcal{P}_1}\mathrm{d}\phi_2.
\end{equation}
Using $\phi_B$, we introduce the period of motion as
\begin{equation}
T=\int_{\phi_2(t=0)}^{\phi_2(t=0)+\phi_B} \frac{C(\phi_2) \mathrm{d}\phi_2}{2(A(\phi_2)\mathcal{P}_2(\phi_2)-B(\phi_2)\mathcal{P}_1)}.
\label{eq:period}
\end{equation}
The $\phi_1$-pendulum completes a full orbit after the time $T$. Within the CISS framework, $T$ defines the time the electron spends inside single-turn molecules. 
Note that for $f=1$, $\theta_1^f=\theta_1^0+\pi C/(T \alpha)$, i.e., $T$ is the only initial-state-dependent parameter that defines how far the electron can move before it loses all of its kinetic energy.

\section{Non-dissipative dynamics with weak SOC}
\label{app:weak_SOC_no_dissipation}

The equations for $\theta_1$ and $\theta_2$ can be written as
\begin{align}
\dot \theta_1=c_1+\frac{\gamma}{2} c_2 f(\delta)-\frac{\gamma}{2} \dot \theta_2 f(\theta_2-\theta_1)-\frac{\gamma}{2} \int_{0}^t \dot \theta_1 \dot \theta_2 \frac{\partial f(\theta_2-\theta_1)}{\partial \theta_2}\mathrm{d}t,\\
\dot \theta_2=c_2+\frac{\gamma}{2\beta} c_1 f(\delta)-\frac{\gamma}{2\beta} \dot \theta_1 f(\theta_2-\theta_1)+\frac{\gamma}{2\beta} \int_{0}^t \dot \theta_1 \dot \theta_2 \frac{\partial f(\theta_2-\theta_1)}{\partial \theta_2} \mathrm{d}t,
\end{align}
where $\delta=\theta_2(t=0)-\theta_1(t=0)$, and $c_i$ are constants determined by the initial conditions, i.e., $c_i=\dot \theta_i(t=0)$.

Let us now assume that the SOC is weak, i.e., $\gamma\to0$.
This means that $\dot \theta_i\simeq c_i$, which leads to the expressions presented in the main text:
\begin{align}
\dot \theta_1\simeq c_1-\frac{\gamma c_2}{2}  (f(c_2 t-c_1 t + \delta)-f(\delta))-\frac{\gamma c_1 c_2}{2(c_2-c_1)} (f(c_2 t-c_1 t + \delta)-f(\delta)),\\
\dot \theta_2\simeq c_2-\frac{\gamma c_1}{2\beta}  (f(c_2 t-c_1 t + \delta)-f(\delta))+\frac{\gamma c_1 c_2}{2\beta (c_2-c_1)} (f(c_2 t-c_1 t + \delta)-f(\delta)).
\end{align}

For the double pendulum, $f(x)=\cos(x)$,
we can easily write the coordinates as well: 
\begin{align}
\theta_1\simeq \theta_1(t=0)+\left(c_1+\frac{\gamma c_2^2 \cos(\delta)}{2(c_2-c_1)}\right) t-\frac{\gamma c_2^2}{2(c_2-c_1)} \frac{\sin(c_2 t-c_1 t + \delta)-\sin(\delta)}{c_2-c_1},\\
\theta_2\simeq \theta_2(t=0)+\left(c_2-\frac{\gamma c_1^2 \cos(\delta)}{2\beta (c_2-c_1)}\right)t+ \frac{\gamma c_1^2}{2\beta(c_2-c_1)} \frac{\sin(c_2 t-c_1 t + \delta)-\sin(\delta)}{c_2-c_1}.
\end{align}

\section{Weak dissipation and SOC}
\label{app:weak_SOC_with_dissipation}

For weak SOC and weak dissipation, we derive
\begin{align}
\dot \theta_1\simeq c_1-\frac{\gamma c_2^2}{2(c_2-c_1)}  (f(c_2 t-c_1 t + \delta)-f(\delta))-\frac{\gamma_1}{2} \theta_1(t)+\frac{\gamma_1}{2}\theta_1(0),\\
\dot \theta_2\simeq c_2+\frac{\gamma c_1^2}{2\beta(c_2-c_1)}  (f(c_2 t-c_1 t + \delta)-f(\delta)).
\end{align}
Note that the equation for $\dot \theta_2$ is the same as for weak SOC (without dissipation). The orbital motion is modified by the presence of dissipation as follows
\begin{equation}
\theta_1=\theta_1^0 e^{-\frac{\alpha t}{2}}+e^{-\frac{\alpha t}{2}}\int_0^t\left[c_1+\frac{\alpha\theta^0_1}{2}-\frac{\gamma c_2^2}{2(c_2-c_1)}(f(c_2 \tau-c_1 \tau + \delta)-f(\delta))\right]e^{\frac{\alpha \tau}{2}}\mathrm{d}\tau.
\end{equation}

\end{appendices}

\bibliography{main}

\end{document}